\documentstyle[12pt]{article}
\font\mybb=msbm10 at 12pt
\def\bb#1{\hbox{\mybb#1}}

\def\IC{{\bb C}}

\begin{document}

\begin{flushright}
IASSNS-HEP-00/35 \\
NSF-ITP-00-36\\
PUPT-1927
\end{flushright}

\begin{center}
{\large{\bf Matrix Theory Interpretation of DLCQ String Worldsheets}}

\bigskip
\bigskip
{\bf G. Grignani $^a$\footnote{Supported by INFN and MURST of
Italy. Permanent Address: Dipartimento di Fisica and Sezione I.N.F.N.,
Universit\`a di Perugia, Via A. Pascoli I-06123, Perugia, Italia.},
P. Orland $^b$\footnote{Supported by PSC-CUNY Research Award Program
Grants nos. 668460, 61508-00, 69466-00, a CUNY Collaborative Incentive
grant 91915-00-06 and the National Science Foundation under Grant
No. PHY94-07194.  Permanent address: Center for Theoretical Physics,
The Graduate School and University Center and Baruch College, The City
University of New York, New York, NY.}, L. D. Paniak
$^c$\footnote{Supported by NSF grant PHY98-02484 and NSERC of Canada.}
and G. W. Semenoff $^a$\footnote{ Supported by the
Ambrose Monell Foundation and NSERC of Canada.  On leave from:
Department of Physics and
Astronomy, University of British Columbia, Vancouver, V6T 1Z1
Canada.}}
\end{center}
\bigskip

\begin{itemize}
\item[a.]Institute for Advanced Study, Einstein Drive, Princeton, NJ 08540.

\item[b.]Institute for Theoretical Physics, University of California, Santa Barbara, CA
93106-4030.

\item[c.]Joseph Henry Laboratories, Department of Physics, Princeton
University, \\Princeton, NJ 08544.

\end{itemize}

\bigskip
\bigskip
\centerline{\bf Abstract}
\bigskip\nonumber

We study the null compactification of 
type-IIA-string perturbation theory at finite temperature.  
We prove a theorem about Riemann surfaces establishing 
that the moduli spaces of infinite-momentum-frame 
superstring worldsheets 
are identical to those of branched-cover instantons in the
matrix-string model conjectured to describe M-theory. 
This means that the identification of string degrees of freedom in the matrix
model proposed by Dijkgraaf, Verlinde and Verlinde is correct and that its 
natural generalization produces the 
moduli space of Riemann surfaces at all orders in the genus expansion.

\thispagestyle{empty}

\newpage
\setcounter{page}{1}

It is widely believed that each of the five known consistent string
theories are limits of a single eleven-dimensional theory called M-theory.
While this theory has not yet been fully mathematically formulated, there
is an interesting proposal, namely the {\it matrix model} \cite{bfss,s}.
This model is conjectured to describe M-theory in a particular kinematical
region, the infinite-momentum frame. In the matrix model, the
superstring is a composite object resembling a necklace of D0-branes.

A nontrivial check of the matrix model proposal would be to use it to
obtain perturbative string theory. There are already convincing arguments
that the non-interacting type-IIA strings emerge
in the appropriate limit of the matrix model \cite{dvv}.  There are
several approaches to deriving superstring interactions along these
lines \cite{dvv}-\cite{brax}. In these, matrix-theory
instantons interpolate between initial states and final states of strings
through a Riemann surface which is a branched cover of the cylinder.
Whether these branched-cover instantons account correctly for
string-scattering amplitudes is an important question.

In this Letter, we show for the 
first time that in the appropriate context, perturbative string theory can
be formulated using {\it only} branched-cover Riemann surfaces.

The matrix model is a formulation of M-theory in the infinite-momentum
frame. Compactification of a spatial direction of M-theory produces type-IIA
string theory. The matrix model becomes
1+1-dimensional, maximally-supersymmetric Yang-Mills theory. 
 
To compare the matrix model directly with infinite-momentum-frame string theory, 
we study the string path integral with a compactified null direction.
The natural quantization of the string in this frame is discrete light-cone 
quantization (DLCQ). 
We find it necessary to introduce a finite temperature by further 
compactifying Euclidean time.
We prove a new theorem on Riemann surfaces, stating
that to any order in the genus expansion, these compactifications restrict
the worldsheets to branched covers of a torus. This differs from the moduli space
of strings without these compactifications, 
which includes {\it all} Riemann surfaces up to conformal diffeomorphisms.

The same branched covers appear in the matrix model. According to Dijkgraaf, Verlinde
and Verlinde \cite{dvv}, the string degrees 
of freedom are simultaneous eigenvalues of the matrices. At finite
temperature, the matrices are defined on a torus \cite{gs} and their eigenvalues, since
they solve polynomial equations, are functions on branched covers of the torus. 
If the matrix model is to agree with perturbative string theory, these branched 
covers must be the full set of Riemann surfaces that contribute to the string path integral.

We compare only the degrees of freedom of the two theories and not the energy spectra. However,
to one-loop order, the spectra are known to coincide \cite{gs}. The coincidence of energy
spectra once higher order corrections are included is still an open question.

The string path integral for the vacuum energy is
\begin{equation}
F=-
\sum_{g,\sigma}g_s^{2g-2} \int [dh_gdXd\Psi]
\exp\left(
-\frac{1}{4\pi\alpha'}\int \sqrt{h}\left(h^{ab}\partial_aX^\mu\partial_b
X^\mu-2\pi i\alpha'\Psi^\mu\gamma\cdot\nabla\Psi^\mu\right)\right)
\label{partf}
\end{equation}
Here, we use, for example, the Neveu-Schwarz-Ramond superstring\footnote{Most 
of our considerations apply to the bosonic
sector of any string theory.  We use the superstring as an example.
For the relationship with matrix theory, however, supersymmetry is
important and that case is more closely related to the Green-Schwarz
superstring.}.  The
string coupling constant is $g_s$ and its powers weight the genus,
$g=0,1,\ldots$, of the string's worldsheet.  There is also a sum over
spin structures, $\sigma$ which, with the appropriate weights, imposes
the GSO projection.  For each value of the genus, $g$, $[dh_g]$ is an
integration measure over all metrics of that genus and is normalized
by dividing out the volume of the worldsheet re-parameterization and
Weyl groups.  We will assume that the metrics of both the worldsheet
and the target spacetime have Euclidean signatures.

We wish to study the situation where the target space has particular
compact dimensions.  Two compactifications will be needed.  The first
compactifies the light-cone in Minkowski space by
making the identification $\frac{1}{\sqrt{2}}\left(t-x^9\right)\sim
\frac{1}{\sqrt{2}}\left(t-x^9\right)+2\pi R$.  In our Euclidean
coordinates it is the identification
\begin{equation}
\left( X^0,\vec X, X^9\right)\sim \left( X^0+\sqrt{2}\pi iR, \vec X,
X^9 -\sqrt{2} \pi R\right)
\label{comp1}
\end{equation}
With this compactification the GSO projection is unmodified.  The
factor of $i$ in the identification of $X^0$ might seem unnatural
since it identifies a real integration variable periodically in a
complex direction.  However, we shall see that, for the path integral
at genus 1, where we can check the result independently by using
operator methods to compute the same partition function, this
identification is indeed the correct thing to do.  We shall postulate
that it also gives the correct partition function at genus
greater than one.

The second compactification that we shall need is that of Euclidean time,
\begin{equation}
\left( X^0,\vec X, X^9\right)\sim \left(X^0+\beta, \vec X, X^9\right)
\label{comp2}
\end{equation}
This compactification, with the appropriate modification of the GSO
projection to make space-time fermions anti-periodic, introduces a
temperature, $T=1/k_B\beta$ where $k_B$ is Boltzmann's constant, so that
(\ref{partf}) computes the thermodynamic free energy.

In order to implement this compactification in the path integral, we
assume that the worldsheet is a Riemann surface $\Sigma_g$ of genus
$g$ whose homology group $H_1(\Sigma_g)$ is generated by the
closed curves,
\begin{eqnarray}&&
a_1,a_2,\ldots, a_g~,~b_1, b_2, \ldots, b_g  \nonumber\\
&&
a_i\cap a_j=\emptyset~,~b_i\cap
b_j=\emptyset~,~ a_i\cap b_j=\delta_{ij}
\label{homology}
\end{eqnarray}
Furthermore, one may pick a basis of holomorphic
differentials $\omega_i\in H^1(\Sigma_g)$ with the properties
\begin{equation}
\oint_{a_i}\omega_j=\delta_{ij}
~~~,~~~
\oint_{b_i}\omega_j=\Omega_{ij}
\label{orthog}
\end{equation}
where $\Omega$ is the period matrix.  It is complex, symmetric,
$\Omega_{ij}=\Omega_{ji}$, and has positive definite imaginary part.

Compactification is implemented by including the possible windings of
the string worldsheet on the compact dimensions.  These form distinct
topological sectors in the path integration in (\ref{partf}).  In the
winding sectors, the bosonic coordinates of the string should have a
multi-valued part which changes by $\beta\cdot$integer or
$(i)\sqrt{2}R\cdot$integer as it is moved along a homology cycle.  The
derivatives of these coordinates should be single-valued functions.
It is convenient to consider their exterior derivatives which can be
expressed as linear combinations of the holomorphic and
anti-holomorphic 1-forms and exact parts,
\begin{equation}
dX^0=\sum_{i=1}^{g}\left( \lambda_i\omega_i+\bar\lambda_i\bar\omega_i\right)+{\rm exact}
~~,~~
dX^9=\sum_{i=1}^{g}\left( \gamma_i\omega_i+\bar\gamma_i\bar\omega_i\right)+{\rm exact}
\label{coord}
\end{equation}
Then, we require
\begin{equation}
\oint_{a_i}dX^0=\beta n_i+\sqrt{2}\pi Ri p_i
~~~,~~~
\oint_{b_i}dX^0=\beta m_i+\sqrt{2}\pi Ri q_i
\end{equation}
\begin{equation}
\oint_{a_i}dX^9=\sqrt{2}\pi R p_i
~~~,~~~
\oint_{b_i}dX^9=\sqrt{2}\pi R q_i
\end{equation}
With (\ref{orthog}), we use these equations to solve for the constants in
(\ref{coord}). With the formula
\begin{equation}
\int\omega_i \bar\omega_j=
\sum_{k=1}^g\left(\oint_{a_k}\omega_i\oint_{b_k}\bar\omega_j-
\oint_{b_k}\omega_i\oint_{a_k}\bar\omega_j\right)=
-2i\left(\Omega_2\right)_{ij}
\end{equation}
we compute the part of the string action which contains the winding integers,
\begin{eqnarray}
S=\frac{\beta^2}{4\pi\alpha'}\left(
n\Omega^{\dagger}-m\right)\Omega_2^{-1} \left( \Omega n-m\right)+ 2\pi
i \frac{ \sqrt{2}\beta R} {4\pi\alpha'}\frac{1}{2}\left[\left(
p\Omega^{\dagger}-q\right)\Omega_2^{-1} \left( \Omega n-m\right)
\right. \nonumber \\ \left.  +\left(
n\Omega^{\dagger}-m\right)\Omega_2^{-1} \left( \Omega
p-q\right)\right]+\ldots
\end{eqnarray}
Note that the integers $p_i$ and $q_i$ appear linearly in a purely
imaginary term in the action. Furthermore, since they come from the
compactification of the light cone, this is the only place that they
will appear in the string path integral (unlike $m_i$ and $n_i$ which
should appear in the weights of the sum over spin structures).  When
the action is exponentiated and summed over $p_i$ and $q_i$, the
result will be periodic Dirac delta functions.  It can be shown that
these delta functions impose a linear constraint on the period matrix
of the worldsheet.  Thus, with the appropriate Jacobian factor, the
net effect is to insert into the path integral measure the following
expression,
\begin{equation}
\sum_{mnrs}e^{-\frac{\beta^2}{4\pi\alpha'}\left(
n\Omega^{\dagger}-m\right)\Omega_2^{-1} \left( \Omega n-m\right)}
\nu^{2g}\left| \det\Omega_2\right| \prod_{j=1}^g
\delta\left(\sum_{i=1}^g\left( n_i+i\nu r_i\right)\Omega_{ij}-\left(
m_j+i\nu s_j\right) \right)
\label{mod}
\end{equation}
where $ \nu=4\pi\alpha'/\sqrt{2}\beta R$ is a fixed constant. 
Consequently, the integration over metrics in the string path
integral is restricted to those for which the period matrix obeys
the constraint
\begin{equation}
\sum_{i=1}^g\left(n_i+i\nu r_i\right)\Omega_{ij}-\left(
m_j+i\nu s_j\right)=0
\label{const}
\end{equation}
for all combinations of the $4g$ integers $m_i,n_i,r_i,s_i$ such 
that $\Omega$ is in a fundamental domain.

Since the columns of the period matrix are linearly independent
vectors, these are $g$ independent complex constraints on the moduli
space of $\Sigma_g$.  Thus its complex dimension $3g-3$ is reduced to
$2g-3$ and there is further discrete data contained in the integers.
One would expect that, when the compactifications are removed, either
$\beta\rightarrow\infty$ or $R\rightarrow\infty$, the discrete data
assembles itself to a ``continuum limit'' which restores the 
complex dimension of moduli space. 

It is interesting to ask whether the Riemann surfaces with the
constraint (\ref{const}) can be classified in a sensible way.  The
answer to this question is yes, a Riemann surface obeys the constraint
(\ref{const}) if and only if it is a branched cover of the torus,
$T^2$, with Teichm\"uller parameter $i\nu$. This is established through the

\vspace{5pt}
\noindent
{\bf Theorem}: $\Sigma_{g}$ is a branched cover of $T^{2}$ if and
only if the period matrix obeys (\ref{const}), for some choice of integers
$m_{i}, n_{i}, r_{i}$ and $s_{i}$. 

\vspace{5pt}
\noindent
{\it Proof}: The generators of the first homology group of $T^2$ are two closed
loops $(\alpha,\beta)$ which span the vector space $H_1(T^2, \IC)$. 
The dual vector space, the first cohomology group, $H^1(T^2, \IC)$ 
is spanned by the basis of holomorphic and anti-holomorphic
differentials $\gamma$ and $\bar\gamma$.  They can be normalized as,
\begin{equation}
\oint_\alpha \gamma=1~~,~~\oint_\beta \gamma=i\nu~~~{\rm and}~~~ \oint_\alpha
\bar\gamma=1~~,~~\oint_\beta\bar\gamma=-i\nu
\end{equation}

The Riemann surface $\Sigma_g$ is a branched cover of $T^2$ if there
exists a continuous, onto, holomorphic map $f$, such that
\begin{equation}
\def\mapright#1{\smash{\mathop{\longrightarrow}\limits^{#1}}}
\Sigma_g ~\mapright{\quad f\quad}~ T^2
\label{map}
\end{equation}
The map $f$ takes closed loops on $\Sigma_g$ to closed loops on $T^2$.  In particular,
the generators (\ref{homology}) must map as
\begin{equation}
\def\mapright#1{\smash{\mathop{\longrightarrow}\limits^{#1}}}
(a_i,b_j)~~\mapright{f} ~~(n_i\alpha+r_i\beta,m_j\alpha+s_j\beta)
\label{maph}
\end{equation}
for some integers $m_i,n_i,r_i,s_i$.
This gives a mapping between the vector spaces $H_1(\Sigma_g,\IC)$ 
and $H_1(T^2,\IC)$.  
A mapping of vector spaces induces a pull-back on the dual vector spaces
\begin{equation}
\def\mapright#1{\smash{\mathop{\longrightarrow}\limits^{#1}}}
H^1(T^2,\IC) ~~\mapright{f^*}  ~~H^1(\Sigma_g,\IC)
\end{equation}
defined by its action on the basis,
\begin{eqnarray}
a_i\circ f^* (\gamma)=\oint_{a_i}f^*(\gamma)
\equiv f(a_i)\circ\gamma=
n_i\oint_\alpha \gamma +r_i\oint_\beta \gamma=n_i +i\nu r_i  \cr
b_j\circ f^* (\gamma )=\oint_{b_j}f^*(\gamma )
\equiv f(b_j)\circ\gamma =
m_j\oint_\alpha \gamma +s_j\oint_\beta 
\gamma = m_j+i\nu s_j
\label{pullback}
\end{eqnarray}

Consider the particular elements of $H_1(\Sigma_g,\IC)$,
\begin{equation}
c_j=\sum_{i=1}^g a_i\Omega_{ij}-b_j
\end{equation}
It can be checked from the definition of the period matrix
(\ref{homology}) (and the fact that it is symmetric) that any
holomorphic differential, $\eta$, on $\Sigma_g$ has the property
\begin{equation}
c_j\circ\eta =
\sum^g_{i=1}\Omega_{ij}\oint_{a_i}\eta -\oint_{b_j}\eta =0
\label{ker}
\end{equation}
The holomorphic nature of the mapping, $f$, guarantees that
$f^*(\gamma)$ is a holomorphic differential on $\Sigma_g$.  Then
it follows that  
\begin{equation}
0=c_j\circ f^*(\gamma)=f(c_j)\circ \gamma=
\sum_{i=1}^g (n_i+i\nu r_i) \Omega_{ij}-(m_j+i\nu s_j)
\end{equation}
which is the constraint on the period matrix in eq.(\ref{const}).

To prove the converse, we must show that if the constraint
(\ref{const}) is satisfied, then a covering map, $f$, exists.  We will
demonstrate this by explicit construction.  Consider the line integral
of a linear combination of the holomorphic differentials on
$\Sigma_g$,
\begin{equation}
z(P)=\int_{P_0}^P \sum_{k=1}^g \lambda_k\omega_k
\label{z}
\end{equation}
with $P_0$ a fixed base-point.  We wish to choose the coefficients
$\lambda_k$ such that this integral defines a map from points
$P\in\Sigma_g$ to the torus $z(P)\in T^2$ whose holomorphic
coordinates are the complex numbers $z(P)$ with the identification
$z\sim z+ p+i\nu q$ where $p$ and $q$ are integers.  The integral
depends on the path of integration.  If the path is changed by a
combination of homology cycles of $\Sigma_g$, $k_ia_i+l_i b_i$, the
integral on the right-hand side of (\ref{z}) changes by
$$
\delta z=\sum_j\lambda_j\left( k_j+\sum_i\Omega_{ji}l_i\right)
$$
$\lambda_k$ must be chosen so that this change is commensurate with
the periods of $T^2$.  This can easily be done if $\Omega$ obeys
(\ref{const}). Then, with the choice $\lambda_i=n_i+i\nu r_i$,
$$
\delta z=\sum_j(n_j+i\nu r_j)k_j+\sum_i(m_i+i\nu s_i)l_i
={\rm integer}+i\nu\cdot{\rm integer}
$$
and we have constructed an explicit covering map, $z(P)$. {\bf Q.E.D.}

As a concrete example, the constraint (\ref{const}) can be solved
explicitly for genus one.
The leading contribution to the free energy of the finite temperature
type II superstring
(without DLCQ) is from the torus amplitude and is given by \cite{aw}
\begin{eqnarray}
\frac{F}{V}=-\int_{\cal F}\frac{d^2\tau}{\tau_2}
\sum_{mn}e^{-\frac{\beta^2 \vert n\tau- m\vert^2}{4\pi\alpha'\tau_2}}
\left(\frac{1}{4\pi^2\alpha'\tau_2}\right)^5
\frac{1}{4\left| \eta(\tau)\right|^{24}}\left[ \left( \theta_2^4\bar\theta_2^4
+\theta_3^4\bar\theta_3^4 + \theta_4^4\bar\theta_4^4\right)(0,\tau)
+
\right.
\nonumber \\
\left.
+e^{i\pi(m+n)}\left( \theta_2^4\bar\theta_4^4+\theta_4^4\bar\theta_2^4
\right)(0,\tau)
-e^{i\pi n}\left( \theta_2^4\bar\theta_3^4+\theta_3^4\bar\theta_2^4
\right) (0,\tau)
- e^{i\pi m}\left( \theta_3^4\bar\theta_4^4 
+ \theta_4^4\bar\theta_3^4\right)(0,\tau)\right]
\label{pf1}
\end{eqnarray}
where $\theta_k(0,\tau)$ are Jacobi theta functions and $\eta(\tau)$
is the Dedekind eta-function.  ${\cal F}$ is the
fundamental domain of the torus,
\begin{equation}{\cal F}~\equiv~\left\{ \tau=\tau_1+i\tau_2 \left|
-\frac{1}{2}<\tau_1\leq\frac{1}{2};~
\vert\tau\vert\geq 1; \tau_2>0 \right. \right\}
\label{fund}
\end{equation}
The modification of this formula by the null compactification can be found using
(\ref{mod}),
\begin{eqnarray}
\frac{F}{V}=-\sum_{\tau\in{\cal F}} \frac{\nu^2}{m^2+\nu^2n^2} 
e^{-\frac{\beta^2 \vert n\tau-m\vert^2}{4\pi\alpha'\tau_2}}
\left(\frac{1}{4\pi^2\alpha'\tau_2}\right)^5
\frac{1}{4\left| \eta(\tau)\right|^{24}}\left[ \left( \theta_2^4\bar\theta_2^4
+\theta_3^4\bar\theta_3^4 + \theta_4^4\bar\theta_4^4\right)(0,\tau)
+
\right.
\nonumber\\
\left.
+e^{i\pi(m+n)}\left( \theta_2^4\bar\theta_4^4+\theta_4^4\bar\theta_2^4
\right)(0,\tau)
-e^{i\pi n}\left( \theta_2^4\bar\theta_3^4+\theta_3^4\bar\theta_2^4
\right) (0,\tau)
- e^{i\pi m}\left( \theta_3^4\bar\theta_4^4 
+ \theta_4^4\bar\theta_3^4\right)(0,\tau)\right]
\label{pf} 
\end{eqnarray}
where the the solution of (\ref{const}) yields the discrete
Teichm\"uller parameter,
$$
\tau=\frac{m+i\nu s}{n+i\nu r}
$$
and one should sum over the integers so that $\tau$ is in the
fundamental domain, ${\cal F}$.

In ref.\cite{gs} it was shown that this formula can also be obtained
by operator methods by finding the spectrum of the non-interacting
type II superstring
in DLCQ and explicitly computing the thermodynamic free energy by summing over
energy states.  This
provides a strong check of the path integral technique for DLCQ we have used
in the present paper.

Modular transformations and identities for theta functions can be used
to rewrite (\ref{pf}) as the Hecke operator \cite{serre}
acting on the partition
function of a superconformal field theory, with torus worldsheet and
target space $R^8$:
\begin{equation}
\frac{F}{V}=-\frac{1}{\sqrt{2}\pi R\beta}{\cal
H}[e^{-\beta/\sqrt{2}R}]*
\left[\left(\frac{1}{4\pi^2\alpha' \tau_2}\right)^4 
\frac{1}{\left| \eta(\tau)\right|^{24}}\left|
\theta_4(0,\tau)\right|^8\right]_{\tau=-1/i\nu}
%\label{pf4}
\end{equation}
A modular transform can be used to write this as
\begin{equation}
\frac{F}{V}=-\frac{1}{\sqrt{2}\pi R\beta}{\cal
H}[e^{-\beta/\sqrt{2}R}]*
\left[\left(\frac{1}{4\pi^2\alpha' \tau_2}\right)^4 
\frac{1}{\left| \eta(\tau)\right|^{24}}\left|
\theta_2(0,\tau)\right|^8\right]_{\tau=i\nu}
\label{pf4}
\end{equation}
The factor in front is the ratio of volumes of $R^8$ and $R^9\times
S^1$ with compactified light cone.  The action of ${\cal H}[p]$ on a
function $\phi(\tau,\bar\tau)$ is defined by
\begin{equation}
{\cal H}[p]*\phi(\tau,\bar\tau)=\sum_{N=0}^\infty p^N
\sum_{\stackrel {kr=N,~r~{\rm odd}} {s~{\rm mod}~k} }
\frac{1}{N}\phi\left(  \frac{r\tau+s}{k}, \frac{ r\bar\tau +s}{k}\right)
\end{equation}

In \cite{gs}, this formula was shown to arise in the $g_s\rightarrow
0$ limit of matrix string theory at finite temperature.  In this
limit, matrix string theory reduces to a theory of eigenvalues of
matrices which naturally live on covers of the torus.  Generally,
these should be branched covers. In the leading order in $g_s$ only
the covers without branch points contribute.  It was shown that the
thermodynamic free energy of the matrix model arising from summing
over unbranched covers is identical to (\ref{pf4}).  The
combinatorics of enumerating them is elegantly accounted for by the
Hecke operator.

In matrix string theory the limit $g_s\rightarrow 0$ corresponds to
large gauge coupling.  This limit projects the theory onto the zeros
of the superpotential.  These zeros occur when all of the matrices are
simultaneously diagonalizable. This is why the matrix model reduces to
a theory of eigenvalues \cite{dvv}.  Beyond the leading order in
strong coupling, the quadratic fluctuations of off-diagonal parts of
the matrices can also be analyzed \cite{bbn}.  It is found that the
fluctuation determinants are ultra-local operators and almost cancel
due to supersymmetry.  Arguments are given that, when carefully
treated, the gauge field sector produces the power of the string
coupling constant $g_s^{2g-2}$ accompanying the branch covers of genus
$g$ (for a discussion of gauge fields, see
refs.\cite{bbn,Billo,Kostov}).  The same argument would apply to an
analysis of the matrix theory at finite temperature.  Even though the
supersymmetry is broken by temperature boundary conditions, since the
fluctuations of non-diagonal fields are governed by
ultra-local operators, the same supersymmetric cancellations should
occur. The remaining theory of diagonal matrices is identical to the
finite temperature Green-Schwarz superstring with worldsheets which
are branched covers of the torus.  In this paper we have shown that
this is exactly what is realized in the DLCQ of string theory at
finite temperature.

What must be done to demonstrate that this limit of matrix string theory 
describes the perturbative type-II superstring? 
It must be shown that the integration
measures over the worldsheets in the string theory and over the
collective coordinates of the branched covers, which can be regarded
as instantons in matrix theory, are identical.

%\section*{References}
The authors thank M. Goresky, J. Harvey, I. Klebanov, J. Koll\'ar, Y. Matsuo,
D. Morrison, E. Verlinde and K. Zarembo for helpful conversations.

\end{document}